\documentclass[prd,aps,twocolumn]{revtex4}
\usepackage[latin1]{inputenc}
\usepackage[english]{babel}
\usepackage{graphicx}
\usepackage{color}
\usepackage{amssymb}
\usepackage{epstopdf}
\usepackage{amssymb}
\usepackage{graphicx}
\DeclareGraphicsExtensions{.pdf,.png,.jpg}
\usepackage{setspace}
\RequirePackage[reqno]{amsmath}
\usepackage{natbib}
\usepackage{color}
\usepackage{mathrsfs}
\usepackage{cancel}
\usepackage{graphics}
\RequirePackage[reqno]{amsmath}

\begin{document}
\title{\bf A monoidal representation for linearized gravity}
% due to the
%
%
\author{Ernesto Contreras\footnote{On leave from Universidad Central de Venezuela} }
\affiliation{Departamento de F\'{\i}sica,
Universidad de los Andes, Apartado A\'ereo {\it 4976}, Bogot\'a, Distrito Capital, Colombia\\
}

\author{Cayetano Di Bartolo}
\affiliation{Departamento de F\'isica, Universidad Sim\'on Bol\'ivar,
Apartado Postal 89000, Caracas 1080-A, Venezuela.}
\author{Lorenzo Leal}
\affiliation{Centro de F\'{\i}sica Te\'orica y Computacional,
Facultad de Ciencias, Universidad Central de Venezuela,
AP 47270, Caracas 1041-A, Venezuela.}
\begin{abstract}
We propose an alternative representation for linear quantum gravity. It is based on the use of a structure that bears some resemblance to 
the Abelian loop representation used in electromagnetism but with the difference that the space of extended objects on which wave functions
take values has the structure of a commutative monoid instead that of Abelian group.
The generator of duality of the theory is realized in this representation and a geometrical interpretation is discussed.

\end{abstract}
\maketitle
\section{Introduction}
It is well known that the introduction of the loop representation 
has opened a new avenue for the quantization of gauge theories such as electromagnetism \cite{dibarto1, contreras1}, 
Yang-Mills \cite{gambi1} and General Relativity \cite{rov1,rov2,baez,barbero,holst,gaul,perez,asht1,asht2,asht3,asht4,asht5,smol}. 
It allows to indroduce an elegant way of solving the constraints of the theories under study as well as, to obtain certain  geometrical information about the 
space of extended objects on which the wave functionals take values.
An example this is given by the Abelian loop representation of the free Maxwell theory (MT).
It is well known that, in this case, the loop representation allows to solve the Gauss constraint immediately \cite{dibarto1, contreras1}. 
Furthermore, despite as
electromagnetism is not a topological theory, it has associated a topological invariant, namely the generator of duality, which
realization in terms of loop-variables leads to a knot invariant, more precisely,
the Gauss linking number \cite{contreras1}.

Within the same spirit, in the context of Abelian field theories, the loop representation has been implemented to quantize the massless Fierz-Pauli theory (FPT) 
\cite{vara, contreras2} with the purpose of explore some geometrical aspects of it. For example, in reference \cite{contreras2} a
representation in terms of skein of loops is obtained. There it was shown that, although the
canonical algebra was fulfilled, a
geometrical interpretation of the physical quantities in term of loops was unclear. Further, the Abelian loop representation is not naturally adapted to the theory in the sense that
tensor indexes of the fields play mixed roles labeling both space coordinates and loops.

In this paper, we consider a monoidal representation as an alternative to deal with the quantization of
symmetric $(0,2)$ Fierz-Pauli tensor fields. As we shall see, it allows to avoid the problem of the mixed indexes because it seems to 
be well-suited to linearized gravity. As a consecuence of that, the wave functionals take values on a space generated by just one kind of extended object
instead of on a list of three tangled Abelian loops as shown in reference \cite{contreras2}.

The organization of the paper is as follows. Section \ref{ab} is devoted to the study of the Abelian loop representation and the subsequent
quantization of both Maxwell and Fierz-Pauli models. In Sec. \ref{mon}, the 
monoidal representation is introduced and 
its application in quantizing the FPT is studied. Finally, we end the paper with some conclusions in Sec. \ref{con}.

\section{Abelian loop representation for Maxwell and Fierz-Pauli models}\label{ab}
The goal of this section is to state the key features for the application of the Abelian loop representation in
the quantization of the MT and the FPT.

We begin by recalling that the Abelian path-space can be described as the set of certain
equivalence classes of curves $\gamma$ in a manifold, which we take as $R^{3}$. The equivalence
relation is given in terms of the so-called form factor $T^{a}(\vec{x},\gamma)$ of the curves
\begin{eqnarray}
T^{a}(\vec{x},\gamma)=\int\limits_{\gamma}dz^{a}\delta^{3}(\vec{z}-\vec{x}),
\end{eqnarray}
as follows: $\gamma$ and $\gamma'$ are said to be equivalent (i.e. they represent the same path), if
their form factor coincide. Closed curves give raise to a subspace of path-space: the
loop-space. It can be seen that the standard composition of curves translates into a
composition of paths that endows path-space with an Abelian group structure. 
Now, we can define the path derivatives $\delta_{a}(\vec{x})$ by
\begin{eqnarray}
u^{a}\delta_{a}(\vec{x})\Psi[\gamma]&:=&\Psi[\gamma\circ u_{\vec{x}}]-\Psi[\gamma]
\end{eqnarray}
where $\circ$ denote the path-space product. The derivative $\delta_{a}(\vec{x})$ measures the
change in the path-dependent wave functional when an infinitesimal path $\delta u$ is
attached to its argument $\gamma$ at the point $\vec{x}$. It is understood that these changes are
considered up to first-order in the infinitesimal vector $u$ associated with the small
path generated by it. As an example of how these operators work, 
we calculate the path-derivative of the form factor. One has
\begin{eqnarray}
T^{a}[\vec{x},\gamma\circ u_{\vec{y}}]&=&\int\limits_{\gamma\circ u_{\vec{y}}}dz^{a}\delta^{3}(\vec{x}-\vec{z})\nonumber\\
&=&T^{a}[\vec{x},\gamma]+u^{b}\delta_{b}^{a}\delta^{3}(\vec{x}-\vec{y})
\end{eqnarray}
Hence,
\begin{eqnarray}\label{dff}
\delta_{a}T^{b}[\vec{x},\gamma]=\delta^{b}_{a}\delta^{3}(\vec{x}-\vec{y}) 
\end{eqnarray}
Now, a geometric representation of a vector field theory arises when the canonical fields
are realized as operators acting onto path-dependent wave functionals $\Psi[\gamma]$. In the MT
the electric and vector potential field operators can be represented in the path-space such as they fulfill the canonical algebra
\begin{eqnarray}\label{conae}
[\hat{A}_{a}(\vec{x},\hat{E}^{b}(\vec{y}))]=i\delta^{b}_{a}\delta^{3}(\vec{x}-\vec{y}) 
\end{eqnarray}
In fact, the realization
\begin{eqnarray}
\hat{E}^{a}|\Psi>&\rightarrow& T^{a}[x,\gamma]\Psi[\gamma]\nonumber\\
\hat{A}_{a}|\Psi>&\rightarrow& i\delta_{a}(\vec{x})\Psi[\gamma] 
\end{eqnarray}
together with equation \eqref{dff}, fulfills the algebra as can be easily verified. 
In the other hand, the Gauss constraint $\partial_{a}\hat{E}^{a}=0$ is identically satisfied if we
deal with closed paths, i.e, if we consider that wave functionals take values on the loop-space.

The program depicted above can be implemented in the Fierz-Pauli model as in reference \cite{contreras2}. In fact, it is immediate to realize 
that the canonical algebra 
\begin{eqnarray}\label{conmutator}
[\hat{h}_{ab}(\vec{x}),\hat{p}^{cd}(\vec{y})]
=\frac{i}{2}(\delta^{c}_{a}\delta^{d}_{b}+\delta^{c}_{b}\delta^{d}_{a})\delta^{3}(\vec{x}-\vec{y}). 
\end{eqnarray}
can be fulfilled with the follow realization
\begin{eqnarray}
\hat{h}_{ab}(\vec{x})|&\rightarrow&\frac{i}{\sqrt{2}}\delta_{a}(\vec{x},\gamma_{b})-\delta_{b}(\vec{x},\gamma_{a})\nonumber\\ 
\hat{p}^{ab}(\vec{x},\gamma)&\rightarrow&\frac{1}{2\sqrt{2}}(T^{a}(\vec{x},\gamma_{b})+T^{b}(\vec{x},\gamma_{a})),
\end{eqnarray}
over wave functionals depending on lists of three closed paths, labeled with the same
indexes used to denote spatial components. Note that, this mixing of space indexes
with ``color'' indexes is crucial for the realization of the algebra.

Unfortanely, this representation for the FPT does not satisfy automatically the
constraints of the theory given by
\begin{eqnarray}\label{constraint}
\partial_{a}\hat{p}^{ab}|\Psi>&=0&\nonumber\\
(\partial_{a}\partial_{b}\hat{h}_{ab}-\nabla^{2}\hat{h})|\Psi>&=&0.
\end{eqnarray}
However, as it is well-known, the linearized constraints state that only the transverse and traceless (TT) components
of the Fierz-Pauli variables are observables in the sense of Dirac. In other words, the TT components of the Fierz-Pauli fields 
fulfills the constraint automatically. For this reason, we must compute the TT part for the canonical pair $(h_{ab},p^{ab})$ making use, for example, of the
TT projector $P_{ab/cd}=P_{ac}P_{db}-\frac{1}{2}P_{ab}P_{cd}$ defined in \cite{contreras2,misner,bar} (here $P_{ab}$ stands for 
the usual transverse projector of electromagentism ), in order to obtain the quantum observables of the theory, i.e,
\begin{eqnarray}\label{projector}
\hat{h}^{TT}_{ab}&=&P_{ab/cd}\hat{h}_{cd}\nonumber\\
\hat{p}^{abTT}&=&P_{ab/cd}\hat{p}^{cd}.
\end{eqnarray}

It is worth mentioning that considering the quantities defined in equation \eqref{projector} as our 
basics quantum operators, the constraints \eqref{constraint} are ``strong'' equalities 
(quantization in the reduced phase space). \\

Before concluding this section, a comparison between Abelian loop representation for Maxwell and Fierz-Pauli models is mandatory.
On one hand, it is plausible to interpret Abelian loops in the MT as quantum Faraday's lines which are closed in absent of sources. 
Note that this fact allows to solve automatically the Gauss constraint. On the other hand, the quantum operator
associated to the electric field can be interpreted as a measure of the density of electric flux and the loop-dependent wave functional depends
only on those features of the loop that are captured by the form-factor, i.e.
$\Psi[\gamma] = \Psi[T^{a}[\vec{x},\gamma]$. However, in the Fierz-Pauli case
such an interpretations are unclear. In the first place,
we are not able to solve immediately the first class constraints by the very choice of closed paths. As was explained above, the constraints are solved
as long as we consider the quantization on the reduced phase space. Secondly, it is not clear if
$p^{abTT}$ can be interpreted as the flux density of \textit{gravitational} Faraday's lines because
it is a function of non-local terms involving the form factor of the loops. In fact, a straighforward calculation reveals that
\begin{widetext}
\begin{eqnarray}
\hat{p}^{abTT}\Psi[\gamma]&=&\frac{1}{\sqrt{2}}(T^{a}[\vec{x},\gamma_{b}]+T^{b}[\vec{x},\gamma_{a}]-\delta_{ab}T^{c}[\vec{x},\gamma_{c}]\nonumber\\
&&-\nabla^{-2}\partial_{c}\partial_{b}T^{a}[\vec{x},\gamma_{c}]-\nabla^{-2}\partial_{c}\partial_{a}T^{b}[\vec{x},\gamma_{c}]+ 
\nabla^{-2}\partial_{a}\partial_{b}T^{c}[\vec{x},\gamma_{c}])\equiv \mathcal{T}^{ab}[\gamma]\Psi[\gamma],
\end{eqnarray}
\end{widetext}
where $\mathcal{T}^{ab}[\gamma]$ is a function of the form factors. For the reasons listed above,
the only we can say about the wave functional is that it take values on a certain space of
\textit{skein} of colored loops, i.e,
\begin{eqnarray*}
\Psi[\gamma]=\Psi[\mathcal{T}^{ab}[\gamma]] 
\end{eqnarray*}

In the next section we shall develope an alternative for the loop representation
that avoid the problem of the space formed by skein of different loops: 
the monoidal representation.

\section{Monoidal representation}\label{mon}
In this section we proceed to construct the monoidal representation for the FPT. 
As a starting point, let us then consider a set of parametrized curves $\gamma$ on $R^{3}$  given by
$\vec{z}_{\gamma}=\vec{z}(l)$, with $l$ the arc length parameter, and define
\begin{eqnarray}\label{definicion}
I^{ab}[\vec{x},\gamma]:=\int\limits_{\gamma}dlu^{a}_{T_{\gamma}}u^{b}_{T_{\gamma}}\delta^{3}(\vec{z}_{\gamma}-\vec{x}), 
\end{eqnarray}
with $\hat{u}_{T_{\gamma}}:=d\vec{z}_{\gamma}/dl$ the tangent vector to $\gamma$ and $a,b=1,2,3$. It is easy to check that \eqref{definicion}
is symmetric and independent of the curve orientation. It is worth mentioning that for any arbitrary parameter $\tau$ we have
\begin{eqnarray}
I^{ab}[\vec{x},\Gamma]=
\int\limits_{\tau_{1}}^{\tau_{2}}d\tau \frac{\dot{z}^{a}\dot{z}^{b}}{|\ \dot{\vec{z}}\ |}\delta^{3}(\vec{x}-\vec{z}(\tau)) 
\end{eqnarray}
where $\dot{z}^{a}:=dz^{a}/d\tau$ and $\tau_{1}<\tau_{2}$ independent of the orientation of $\gamma$.

Let $\mathcal{M}$ be the space which elements $\Gamma$ are the union of disjoint curves, i.e,
$\Gamma=\gamma_{1}\cup....\cup\gamma_{n}$ with arbitrary $n$. From the definition \eqref{definicion}, it follows that 
\begin{eqnarray}
I^{ab}[\vec{x},\Gamma]&=&I^{ab}[\vec{x},\gamma_{1}\cup...\cup\gamma_{n}]\nonumber\\
                      &=&\sum\limits_{i=1}^{n}I^{ab}[\vec{x},\gamma_{i}]. 
\end{eqnarray}

We shall say that two curves $\Gamma$ and $\Gamma'$ represent the same element or \textit{path} 
if $I^{ab}[\vec{x},\Gamma]=I^{ab}[\vec{x},\Gamma']$. The space formed by such a paths will be denoted by $\mathfrak{M}$  .

For elements in $\mathfrak{M}$ we shall define the product $\circ$ between two paths $\Gamma_{1}$ and $\Gamma_{2}$ as follows
\begin{eqnarray}\label{multipli}
I^{ab}[\vec{x},\Gamma_{1}\circ\Gamma_{2}]&=&I^{ab}[\vec{x},\Gamma_{2}\circ\Gamma_{1}]\nonumber\\
&:=&I^{ab}[\vec{x},\Gamma_{1}\cup\Gamma_{2}]\nonumber\\ 
&=&I^{ab}[\vec{x},\Gamma_{1}]+I^{ab}[\vec{x},\Gamma_{1}].
\end{eqnarray}
Note, from the above definition, that this product is commutative. Furtheremore, 
for an element $\Gamma_{e}\in\mathfrak{M}$ such as $I^{ab}[\vec{x},\Gamma_{e}]=0$ 
(the null path)
occurs that
\begin{eqnarray}
I^{ab}[\vec{x},\Gamma\circ\Gamma_{e}]=I^{ab}[\vec{x},\Gamma].
\end{eqnarray}
In this manner, $\Gamma_{e}$ is the neutral element for the multiplication defined in \eqref{multipli}. In summary,
the elements of $\mathfrak{M}$ along with relation \eqref{multipli} form a commutative monoid, i.e, a semigroup with an identity.

Let us now consider square integrable functionals (at least formally) $\Psi:\mathfrak{M}\rightarrow C$. We define the path derivative operator
$\delta_{cd}(\vec{y},\hat{v})$ as
\begin{eqnarray}\label{defderivada}
v^{c}v^{d}\delta_{cd}(\vec{y},\hat{v})\Psi
:=\lim\limits_{L\rightarrow0}\frac{\Psi[\Gamma\circ\Gamma[\vec{y},\hat{v}]]-\Psi[\Gamma]}{L}. 
\end{eqnarray}

As can be seen, this derivative computes the change of $\Psi$ when an infinitesimal path  $\Gamma[\vec{y},\hat{v}]$ starting at $\vec{y}$ 
in the $\hat{v}$ direction,
is appended to the path $\Gamma$. Now, with the above definition, is straightforward to prove that the path derivative for $I^{ab}[\vec{x},\Gamma]$ is given by
\begin{eqnarray}\label{derivative}
\delta_{cd}(\vec{y},\hat{v})I^{ab}[\vec{x},\Gamma]=\frac{1}{2}(\delta^{a}_{c}\delta^{b}_{d}+\delta^{a}_{d}\delta^{b}_{c})\delta^{3}(\vec{y}-\vec{x}) 
\end{eqnarray}

With these tools at hand we are ready to obtain a monoidal representation for linearized gravity.
In order to carry this out, let us consider the realization
\begin{eqnarray}\label{rep}
\hat{h}_{ab}\Psi[\Gamma]&\rightarrow& i\delta_{ab}(\vec{x},\hat{v})\Psi[\Gamma]\nonumber\\
\hat{p}^{ab}\Psi[\Gamma]&\rightarrow& I^{ab}[\vec{x},\Gamma]\Psi[\Gamma],
\end{eqnarray}
which fulfills the canonical algebra (see Eq. \eqref{conmutator}). 
Note that in the above realization, we have chosen the following ``polarization'' for the
quantum operators: the momenta are diagonal, while the fields act by taking (path)
derivatives. This choice, inspired by the Maxwell case, is not mandatory, and the
roles of momenta and fields could be interchanged (with an appropriate change
of sign). It is also noticeable that this prescription is naturally adapted to linearized gravity in the same way that 
the Abelian loop representation in the electromagnetic case, in the sense that no further intrepretation for the tensor indexes $a,b$ is needed. In the same way,
the wave functional depends on a non-trivial \textit{tangle} of the same closed path $\Gamma$
instead of skein of colored loops (as it was shown in the previous section). Unfortunately, the first class constraints cannot be realized in this case
either and therefore we must carry out the quantization in the reduced phase space as in the FPT case. In spite of that, 
the monoidal representation can be used to obtain an interesting geometric interpretation of the generator of duality.
With this purpose, let us first recall how the generator arises in the theory  and subsequently we shall proceed to 
quantize it in terms of monoid variables. 

It is well known that,
analogously 
to what occurs in the FMT, the Fierz-Pauli model shows invariance of the 
equations of motion under the transformations
\begin{eqnarray}
p^{abTT}&\rightarrow&\frac{1}{2}(\mathcal{O}h^{TT})_{ab}\nonumber\\
\frac{1}{2}(\mathcal{O}h^{TT})_{ab}&\rightarrow&-p^{abTT},
\end{eqnarray}
where $(\mathcal{O}h)_{ab}=\frac{1}{2}\varepsilon^{acd}\partial_{c}h_{db} + \varepsilon^{bcd}\partial_{c}h_{da}$ is the symmetrized curl.
As a guide of the comparison with the Maxwell case, it is useful to keep in mind
that $p^{abTT}$ is the analogous to the transverse part of the electric field, while $h^{TT}_{ab}$
would relate with the transverse part of the vector potential and its symmetrized
curl $(\mathcal{O}h^{T T} )_{ab}$ is analogous to the magnetic field. Infinitesimally, the duality transformations can be written down as
\begin{eqnarray}\label{dt}
\delta p^{abTT}&=&\frac{\theta}{2}(\mathcal{O}h^{TT})_{ab}\nonumber\\
\frac{1}{2}\delta h_{ab}^{TT}&=&\theta\nabla^{-2}(\mathcal{O}p^{TT})^{ab}
\end{eqnarray}
which, as in the Maxwell theory, correspond to a SO(2) rotation. A straightforward
application of Noether's theorem leads to the conserved quantity which generates
the infinitesimal duality rotations (see reference \cite{contreras2} for details). Now, in the monoidal 
representation this generator of duality can be shown to be given by 
\begin{widetext}
\begin{eqnarray}\label{godm}
\hat{G}&=&\int d^{3}x\delta_{ab}(\vec{x},\hat{v})\varepsilon^{acd}\partial_{c}\delta_{db}(\vec{x},\hat{v})
  +\int d^{3}x\partial_{b}\delta_{ab}(\vec{x},\hat{v})\varepsilon^{acd}\partial_{c}\nabla^{-2}\partial_{e}\delta_{de}(\vec{x},\hat{v})\nonumber\\
&+&\int d^{3}xI^{ab}[\vec{x},\Gamma]\varepsilon^{acd}\nabla^{-2}\partial_{c}I^{db}[\vec{x},\Gamma]
   +\int d^{3}x\partial_{b}I^{ab}[\vec{x},\Gamma]\varepsilon^{acd}\partial_{c}\nabla^{-4}\partial_{e}I^{de}[\vec{x},\Gamma].
\end{eqnarray}
\end{widetext}

It is worth to mentioning that the expression \eqref{godm} have formal similarities with generator of duality of the MT reported in reference
\cite{contreras1}. One of these features is that both are written in term of products which involves either path 
derivatives or ``form factors''. Another one
is that non-local terms appear as a consecuence of the inverse Laplacian operator $\nabla^{-2}$ which
allowed to define the infinitesimal transformation in equation \eqref{dt}. Finally, only the 
``form factor" part of the generator 
can be interpreted in terms of features of the space of extended objects\footnote{The path derivative only has a geometric meaning 
in the dual representation where $\hat{h}^{TT}_{ab}$ is diagonal}. In order to show this, 
let us concentrate in the third term of the generator of duality\footnote{It can be shown that the fourth term has not a simple interpretation}
 which after some manipulation can be written as
\begin{widetext}
\begin{eqnarray}\label{tres}
\hat{G}_{3}\propto -\frac{1}{4\pi}\oint\limits_{\Gamma_{1}}dl_{\Gamma_{2}}\oint\limits_{\Gamma_{2}}dl_{\Gamma_{2}}(\hat{u}_{T_{\Gamma_{1}}}\cdot\hat{u}_{T_{\Gamma_{2}}})
\int\limits_{\gamma}^{\vec{z}_{\Gamma_{1}}}dl_{\gamma}(\hat{u}_{T_{\Gamma_{1}}}\times\hat{u}_{T_{\Gamma_{2}}})\cdot\hat{u}_{T_{\gamma}}
\delta^{3}(\vec{z}_{\Gamma_{2}}-\vec{w}_{\gamma}).
\end{eqnarray}
\end{widetext}

In the former expression, it is assumed that each line integrals have different supports, namely $\Gamma_{1}$ and $\Gamma_{2}$ (actually
both integrals are defined on the same path $\Gamma$ but, as we shall see, our choice of different paths will simplify the geometrical interpretation). 
Furthermore, $\gamma_{1}$ is an auxiliar infinte straight curve that does not belong to the monoide space $\mathfrak{M}$ but it
has been introduced for convenience (see for example reference \cite{contreras1} for details). In the other hand,
unitary tangent vectors $\hat{u}_{T_{\Gamma_{1}}}$ and $\hat{u}_{T_{\Gamma_{2}}}$
belong to the closed paths $\Gamma_{1}$ and $\Gamma_{2}$
respectively. In contrast, $\hat{u}_{T_{\gamma}}$ is tangent
to the open straight curve which starts in the spatial infinity and
ends on $\Gamma_{2}$. Now, the triple product $(\hat{u}_{T_{\Gamma_{1}}}\times\hat{u}_{T_{\Gamma_{2}}})\cdot\hat{u}_{T_{\gamma}}$ 
force the vectors $\hat{u}_{T_{\Gamma_{1}}}$,
$\hat{u}_{T_{\Gamma_{2}}}$ and $\hat{u}_{T_{\Gamma_{2}}}$ to construct a non-degenerated
volume if we want to detect non-vanishing contributions. We can see that these curves $\Gamma_{1}$ and
$\Gamma_{2}$ are linked together
by the fact that $\Gamma_{1}$ is obligated to intersect once the
open curve, and if in such intersection the
three vectors involved form a non-degenerate figure-volume, then
this will ``count'' out one contribution. 

In base on the former discussion, we can realize the intriguing similarity between $\hat{G}_{3}$ and the Gauss Linking Number (GLN). The only difference
with the GLN is that $\hat{G}_{3}$ contain also the product $\hat{u}_{T_{\Gamma_{1}}}\cdot\hat{u}_{T_{\Gamma_{2}}}$ which is a 
measure of the angle between the tangent vectors of the closed paths. In this sense we can 
say that this term will contribute to the generator
as long as tangent vectors $\hat{u}_{T_{\Gamma_{1}}},\hat{u}_{T_{\Gamma_{2}}},\hat{u}_{T_{\gamma}}$
form a trihedron but also the dihedral angle between planes intersecting on the line generated by $\hat{u}_{T_{\gamma}}$ must be neceseraly different 
from $0$ and $\pi/2$. However, instead 
of obtain $\pm1$ every time this occurs (as for the Gauss Linking Number) our result is angle dependent, or in other words, is metric dependent. 
This fact was expected 
because the term from where it was derived, namely, $I^{ab}[\vec{x},\Gamma]\varepsilon^{acd}\partial_{c}I^{db}[\vec{x},\Gamma]$ is
metric dependent (although at the first glance, looks like a Chern-Simon term). Note that a similar result was obtained for the generator of duality for the 
MT (see reference \cite{contreras1}). In fact, the electric part of the generator can be written as 
\begin{eqnarray}\label{emgod}
\oint\limits_{\Gamma_{1}}dl_{\Gamma_{2}}\oint\limits_{\Gamma_{2}}dl_{\Gamma_{2}}
\int\limits_{\gamma}^{\vec{z}_{\Gamma_{1}}}dl_{\gamma}(\hat{u}_{T_{\Gamma_{1}}}\times\hat{u}_{T_{\Gamma_{2}}})\cdot\hat{u}_{T_{\gamma}}
\delta^{3}(\vec{z}_{\Gamma_{2}}-\vec{w}_{\gamma}),\nonumber\\
\end{eqnarray}
which, in absence of the scalar product $\hat{u}_{T_{\Gamma_{1}}}\cdot\hat{u}_{T_{\Gamma_{2}}}$, it
is an analytical expression of the Gauss Linking number between Abelian loops.

\section{Concluding remarks}\label{con}
In this work, a well-suited geometrical representation for the massless Fierz-Pauli theory has been obtained. It presents 
advantages when is compared with previous attempts. For instance it allows to represent symmetric quantum operators 
in a way that a misleading interpretation of tensor indexes of the theory is avoided. As a consecuence, the space
on which wave funtionals take values is expanded by monoidal closed paths instead of on skein of colored Abelian loops.
Moreover, the monoidal representation developed here, allows to carry out a kind of geometric interpretation of
interesting objects like the generator of duality of the theory. Although a knot invariant is not obtained, the resemblance
with the Gauss Linking Number is intriguing and further research along this line is mandatory. If the monoidal 
representation allows to obtain some hints that can be further applied or generalized to  
full gravity or if it can provide more illumination respect to the standard loop formulation is under investigation.

\section{Acknowledgments}
E.C acknowledges Pedro Bargue\~no for his constructive and valuable suggestions.

\end{document}